\documentclass[]{interact}

\usepackage[compatible]{algpseudocode}
\usepackage{algorithm}
\usepackage{multirow}
\usepackage{epstopdf}% To incorporate .eps illustrations using PDFLaTeX, etc.
\usepackage[caption=false]{subfig}% Support for small, `sub' figures and tables

\usepackage[numbers,sort&compress]{natbib}% Citation support using natbib.sty
\bibpunct[, ]{[}{]}{,}{n}{,}{,}% Citation support using natbib.sty
\theoremstyle{plain}% Theorem-like structures provided by amsthm.sty

\theoremstyle{definition}

\theoremstyle{remark}

\begin{document}

%\articletype{ARTICLE TEMPLATE}% Specify the article type or omit as appropriate

\title{GLS and VNS Based Heuristics for Conflict-Free Minimum-Latency Aggregation Scheduling in WSN}

\author{
\name{Roman Plotnikov\textsuperscript{a}\thanks{CONTACT Roman Plotnikov. Email: prv@math.nsc.ru},
Adil Erzin\textsuperscript{a,b},
and Vyacheslav Zalyubovskiy\textsuperscript{a}}
\affil{\textsuperscript{a}Sobolev Institute of Mathematics, Novosibirsk, Russia; \textsuperscript{b}Novosibirsk State University, Novosibirsk, Russia}
}

\maketitle

\begin{abstract}
We consider a conflict-free minimum latency data aggregation problem that occurs in different wireless networks. Given a network that is presented as an undirected graph with one selected vertex (a sink), the goal is to find a spanning aggregation tree rooted in the sink and to define a conflict-free aggregation minimum length schedule along the arcs of the tree directed to the sink. Herewith, at the same time slot, each element of the network can either send or receive at most one message. Only one message should be sent by each network element during the whole aggregation session, and the conflicts caused by signal interference should be excluded. This problem is NP-hard and remains NP-hard even in the case when the aggregation tree is given. Therefore, the development of efficient approximate algorithms is very essential for this problem. In this paper, we present new heuristic algorithms based on the genetic local search and the variable neighborhood search metaheuristics. We conducted an extensive simulation that demonstrates the superiority of our algorithms compared with the best of the previous approaches.
\end{abstract}

\begin{keywords}
Wireless sensor networks; data aggregation; minimum latency; variable neighborhood search; memetic algorithm; simulation
\end{keywords}

\section{Introduction}

One of the most common applications of wireless sensor networks (WSNs) is the collection of sensing information to a designated node called a sink \cite{Bagaa14}. This kind of all-to-one communication pattern is also known as {\it convergecast}. Since sensor nodes are equipped with radio transmitters with limited transmitting range, hop-by-hop communications are usually used to deliver the information from sensor nodes to the sink. WSN is commonly modeled as a graph, where vertices represent sensor nodes, and any two nodes are connected if the distance between them is within their transmission range. The convergecast process, in this case, is modeled by building a logical tree on top of the physical topology with the sink located at the root assuming that packets are routing along the tree's arcs.

Since energy consumption is the most critical issue of WSNs application, energy efficiency becomes one of the primary design goals for a convergecast protocol. Obviously, the convergecast of all raw data will cause to a burst traffic load. To reduce the traffic load, in-network data aggregation can be applied. Here aggregation referees to the process when the relay nodes merge the received data with their own data by means of data compression, data fusion or aggregation function \cite{Malhotra11}. In this case, each sensor node has to transmit only once and the transmission links form a tree, which is called the {\it aggregation tree}.

When a node sends the data to the receiver, a {\it collision} or {\it interference} can occur at the receiver if the transmission interferes with signals concurrently sent by other nodes, and thus the data should be retransmitted. Since the retransmissions cause both extra energy consumption and an increase of convergence time, protocols able to eliminate the collisions are on demand. A common approach is to assign a sending timeslot to each node in such a way that all data can be aggregated without any collision on their way to the sink node, which is known in the literature as {\it time division multiple access} (TDMA)-based scheduling. Most of the scheduling algorithms adopt the {\it protocol interference model}  \cite{Hromkovic96}, which enables the use of simple graph-based scheduling schemes. The {\it physical model} is based on the signal-to-interference-plus-noise-ratio (SINR) and provides a better solution in terms of realistic capturing of interference from multiple transmissions \cite{Gupta00}.

In terms of common objectives of TDMA-based scheduling algorithms, the following two are most fundamental with respect to data aggregation in WSNs: minimizing schedule length or latency and minimizing energy consumption. In terms of design assumptions, the algorithms differ mainly in the following categories: use of communication and interference models, centralized or distributed implementation, topology assumption, and types of data collection \cite{Incel12}. In this paper, we consider the problem of minimization aggregation latency assuming the collision-free transmission under the protocol model and uniform transmission range. This problem is known as the Minimum-Latency Aggregation Scheduling (MLAS) \cite{Xu11} or Aggregation Convergecast Scheduling \cite{Pan16}.

In terms of computational complexity, the MLAS problem is NP-hard \cite{Chen05}, so most of the existing results in the literature are heuristic algorithms, that are usually comprised of two independent phases: construction of an aggregation tree and link transmission scheduling. It is worth mentioning that both these problems are very hard to solve. On one hand, there is no result describing the structure of an optimal aggregation tree for a given graph. On the other hand, even on a given aggregation tree, optimal time slot allocation is still NP-hard \cite{erPyat16}.

To overcome the above-mentioned problems, in this paper, we propose two new heuristic algorithms that reduce the aggregation delay by constructing proper a suboptimal aggregation tree. The first algorithm is based on genetic algorithm and uses local search procedures together with random mutation operator. The second one is based on the variable neighborhood search metaheuristic. Different aggregation trees are generated and evaluated within the both algorithms that gives more chances to find better solutions comparing with the state-of-the-art approaches. We conduct extensive simulation experiments to demonstrate the quality of the solutions provided by the proposed methods vs. Integer Programming based optimal algorithm and the the best of the known heuristics.

This paper is an extension of the paper \cite{optima18}. As an addition material, this paper contains a new VNS based algorithm. Besides, to make our results reproducible, for this paper, we launched the numerical experiment on the public available test instances instead of the ones generated by ourselves.

The rest of this paper is organized as follows: Section 2 reviews the related work. In Section 3, we provide assumptions and formulation of the problem. A new heuristic algorithm based on genetic algorithm and local search metaheuristics is presented in Section 4. A new VSN based heuristic algorithm is presented in Section 5. Section 6 contains results and analysis of an experimental study, and Section 7 concludes the paper.

\section{Related Work}
In one of the early works on MLAS, Chen et al. \cite{Chen05} consider a slightly generalized version of the MLAS, Minimum Data Aggregation Time (MDAT) problem, where only a subset of nodes generates data. They proved that MDAT is NP-hard  and designed an approximation algorithm with guaranteed performance ratio $\Delta -1$, where $\Delta$ is the maximal number of sensor nodes within the transmission range of any sensor. In \cite{Zhu18} the authors propose an approximation algorithm with guaranteed performance ratio $\frac{7\Delta}{\log_2|S|}+c$, where $S$ is the set of sensors containing source data, and $c$ is a constant.

As mentioned before, most MLAS algorithms solve the problem in two consecutive phases: aggregation tree construction and link scheduling. Shortest Path Tree (SPT) and Connected Dominating Set (CDS) are the usual patterns for the aggregation tree. In SPT based algorithms \cite{Malhotra11}, \cite{Incel12}, \cite{Chen05}, a sensor transmits data through a path with minimum length, which reduces aggregation delay, but they do not take potential collisions into consideration. As for CDS based algorithms, due to the topological properties of CDS, it is often possible to prove for them upper bound of data aggregation delay, which usually depends on network radius $R$ and maximum node degree $\Delta$. Huang et al. \cite{Huang07} proposed an aggregation scheduling method based on CDS with the latency bound $23R+\Delta -18$. Based on the deeper study of the properties of neighboring dominators in CDS, Nguyen et al. \cite{Nguyen11} provided a proof of an upper bound $12R+\Delta - 12$ for their algorithm.

Despite the ability to have a delay upper bound for CDS based algorithms, the upper bounds are much greater than their real performance shows. Moreover, a dominating node is likely to be a node of large degree, which may have negative effect on aggregation delay. SPT has similar problems for the sink node, which takes all its neighbors as its children. It was shown in \cite{Tian11} that an optimal solution could be neither SPT nor CDS based.   De Souza et al. \cite{Souza13} constructed an aggregation tree by combining an SPT and a minimum interference tree built by Edmond's algorithm \cite{Edmonds67}. In \cite{Pan16}, the authors proposed a Minimum Lower bound Spanning Tree (MLST) algorithm for aggregation tree construction. To achieve a small delay lower bound, they use the sum of the receivers' depth and child number as the cost of the transmission link. However, the problem of finding the optimal aggregation tree for the MLAS problem remains unsolved.

Genetic algorithm \cite{Sivanandam08} (GA) is one of the most common metaheuristics that is used for the approximate solution of NP-hard discrete optimization problems including those in the WSN domain. In particular, in \cite{Gruber06,Hussain2007,Sudha2011,Liu2011}, different GA based approaches were used to solve problems associated with the minimization of energy consumption. In addition, researchers in \cite{Yen11} proposed GAs for the multiple QoS (quality of service) parameter multicast routing problem in a Mobile ad hoc network (MANET). In \cite{Nanhao13} a GA based framework was presented for the custom performance metric optimization of WSN. In \cite{ploErZal17} a GA based algorithm was proposed to solve the convergecast scheduling problem with an unbounded number of channels, where only conflicts between the transmitters to the same addressee were taken into account.

A variable neighborhood search (VNS) is a metaheuristic approach proposed by Hansen and Mladenovic \cite{Hansen01}. The basic idea of VNS is to locally explore better solutions based on a dynamic neighborhood model.
VNS-based heuristics are applied to different combinatorial optimization problems including those in the WSN domain.
In \cite{Guney09}, the authors jointly solve the point coverage problem, sink location problem and data routing problem on heterogeneous sensor networks. Su et al. \cite{Su15} proposed the VNS heuristic to minimize transmission tardiness in data aggregation scheduling.  Plotnikov et al. \cite{Plotnikov16} investigated problem of finding an optimal communication subgraph in a given edge-weighted graph, which translates to the problem of minimization transmission energy consumption in a WSN.

\section{Problem Formulation}

We consider a WSN consisting of stationary sensor nodes with one sink. All sensors are homogeneous. We use a protocol interference model \cite{Hromkovic96}, which is a graph theoretic approach that assumes correct reception of a message if and only if there is no simultaneous transmissions within proximity of the receiver. For simplicity, we assume that the interference range is equal to the transmission range, which is the same for each sensor. Then the WSN with sink node $s$ can be represented as a graph $G = (V, E)$, where $V$ denotes all the sensor nodes and $s \in V$. An edge $(u, v)$ belongs to $E$ iff the distance between the nodes $u$ and $v$ is within the transmission range. We also assume that time is divided in equal-length slots under the assumption that each slot is long enough to send or receive one packet.
The problem considered in this paper is defined as follows: Given a connected undirected graph $G = (V, E)$, $|V| = n$ and a sink node $s \in V$, find the minimum length schedule of data aggregation from all the vertices of $V \setminus \{s\}$ to $s$ (i.e., assign a sending time slot and a recipient to each vertex) under the following conditions:

\begin{itemize}
 \item each vertex sends a message only once during the aggregation session (except the sink which always can only receive messages);
 \item once a vertex sends a message, it can no longer be a destination of any transmission;
 \item if some vertex sends a message, then during the same time slot none of the other vertices within a receiver's interference range can send a message;
 \item a vertex cannot receive and transmit at the same time slot.
\end{itemize}

As it follows from this formulation, the data aggregation have to be performed along the directed edges (arcs) of a spanning tree rooted in $s$. Since it is convenient to consider the arcs when constructing an aggregation tree, we also introduce a directed graph $G_{dir} = (V, A)$ constructed  from $G$ by replacing each edge with two oppositely directed arcs and excluding the arcs starting from $s$.

\section{Genetic Local Search}\label{sec:heuristics}
As was mentioned before, there are a lot of known heuristic approaches for the approximate solution to the considered problem. As a rule, each of these methods consists of two stages. At the first stage, an aggregation tree is built, and at the second stage, a conflict-free schedule is constructed. We suggest a new approach where different aggregation trees are examined within an algorithm, based on a GA metaheuristic, that applies a local search procedure to the offspring as well as a fully randomized mutation procedure. Such an algorithm is often called \emph{memetic} or \emph{genetic local search} (GLS).

Since the considered problem remains NP-hard even for a given aggregation tree \cite{erPyat16}, the problem of calculating the difference between the values of minimum length aggregation schedules of two aggregation trees is intractable too. Therefore, any type of local search is not applicable, within the acceptable time, to this problem. Instead, we propose a local search procedure for the reduced problem, where only the \emph{primary conflicts} (i.e., the conflicts that occur on attempt of simultaneous sending messages from different transmitters to the same recipient) are taken into account. In this case, the difference of the objective values between two neighboring aggregation trees may be calculated efficiently. Additionally, after applying a local search procedure, an approximate scheduling algorithm for the initial problem can be applied to the aggregation tree. As an approximate algorithm of calculation of a conflict-free schedule for a given aggregation tree, we use the heuristic algorithm Neighbor Degree Ranking (NDR) \cite{Pan16}. Note that this method can change the aggregation tree because of the Supplementary Scheduling subroutine.

A brief description of the main steps of the proposed algorithm is presented in Algorithm \ref{fig:gls}. Like other GA based methods, this algorithm imitates an evolutionary process. Once created  by the Initialization procedure \emph{population}, a set of feasible solutions to the considered problem, is iteratively updated within the \textbf{while} loop. In each iteration of this loop, the current population generates an offspring after applying Selection and Crossover procedures, and then the elements of the offspring are modified by Mutation and LocalSearch procedures. Then, the fitness of each offspring element is calculated, and inside the Join procedure the fittest elements are included in the next generation.

\begin{algorithm}[!hbtp]
\begin{algorithmic}[1]
\STATE \emph{Input}: $G_{dir} = (V, A)$ is a communication graph, $PopSize$, $OffspSize$, $FPItCount$, $SPProportion$, $PM$, $PLS$, $K_{max}$ --- algorithm parameters;
\STATE \emph{Output}: $T$ --- spanning tree in $G$ rooted in $s$;
\STATE Initialization;
\STATE FitnessCalculation(population);
\WHILE {stop condition is not met}
\STATE Selection;
\STATE Crossover;
\STATE Mutation;
\STATE LocalSearch;
\STATE FitnessCalculation(offspring);
\STATE Join;
\STATE Let $T$ be the best tree among the current population;
\ENDWHILE
\end{algorithmic}
\caption{Genetic local search} \label{fig:gls}
\end{algorithm}

The algorithm takes as an input a communication graph  $G_{dir}$ and the following parameters:

\begin{itemize}
  \item $PopSize$ --- the size of population;
  \item	$OffspSize$ --- the size of offspring;
  \item	$FPItCount$ --- the number of iterations in the first population construction procedure;
  \item	$SPProportion$ --- the ratio of shortest-path trees in the starting population;
  \item	$PM$ --- the probability of mutation;
  \item	$PLS$ --- the probability of a local search.
  \item	$K_{max}$ --- the maximum possible number of iterations in the mutation procedure
\end{itemize}

The main steps are described in detail in the next subsections.

\subsection{Initialization}
The first population is generated within the Initialization procedure. At first, we generate three trees with the most efficient known heuristics: a shortest-path tree (e.g., constructed by the Dijkstra algorithm), a tree constructed by the the Round Heuristic (RH) \cite{Beier00}, that appears to be very effective for the simplified problem with only the primary conflicts, and Minimum Latency Spanning Tree (MLST) introduced in \cite{Pan16}. After the shortest-path tree is constructed, the length of the shortest path from each vertex to the sink is known. Let $l(v)$ be the length (number of edges) of the shortest path from vertex $v \in V$ to $s$. The next trees added to the population are generated by two procedures: $RandomShortestPath$ and $RandomMinDegree$. The procedure $RandomShortestPath$ starts with a tree $T = (\{s\}, \emptyset)$, and then for each vertex $v \in V \setminus \{s\}$ an arc from a set $A_v = \{ (u, v) | (u,v) \in A, l(u) = l(v) - 1\}$ is chosen at random and added to the current tree. The procedure $RandomMinDegree$ starts with a tree $T = (\{s\}, \emptyset)$ as well, and then
an arc from $A$ that connects a vertex from the current tree with a vertex from $V$
that does not belong to the current tree is sequentially chosen at random and added to the
current tree, and the probability of an arc choice is inversely proportional to the degree of a corresponding vertex in the current tree. A new tree is added to the population only if it is not a copy of an existing one. The Initialization step requires three parameters: $PopSize$ –-- the maximum size of the population, $SPProportion$ --- an approximate part of the trees generated by the procedure $RandomShortestPath$, and $FPItCount$ --- the maximum number of successive attempts to generate a tree. The pseudocode of the Initialization procedure can be found in Algorithm. \ref{fig:gls_init}.

\begin{algorithm}[!hbtp]
\begin{algorithmic}[1]
\STATE \emph{Input}: $G_{dir} = (V, A)$ --- a communication graph, $PopSize$, $FPItCount$, $SPProportion$ --- algorithm parameters;
\STATE \emph{Output}: $P$ --- population (a set of spanning trees in $G$ rooted in $s$);
\STATE	Set $P \leftarrow \emptyset$, $i \leftarrow 0$;
\STATE	Add the trees constructed by Dijkstra algorithm, Round Heuristic, and Minimum Latency Spanning Tree to $P$ ;
\WHILE {$i < FPItCount$ and $|P| < PopSize$}
\STATE	  $x$ $\leftarrow$ random real value between 0 and 1
\IF {$x < SPProportion$}
\STATE Set $T \leftarrow RandomShortestPath()$;
\ELSE
\STATE Set $T \leftarrow RandomMinDegree()$;
\ENDIF
\IF {$P$ does not contain $T$}
\STATE Add $T$ to $P$;
\ENDIF
\STATE Set $i \leftarrow i+1$;
\ENDWHILE
\end{algorithmic}
\caption{Initialization} \label{fig:gls_init}
\end{algorithm}

\subsection{Fitness calculation}
In order to estimate the quality of every tree in the population its fitness should be calculated. Fitness is a positive value which is higher when the value of the objective function is closer to optimal. Let $L(T)$ be the length of an aggregation schedule for a spanning tree $T$. Then the fitness is $1 / L(T)$. As was mentioned before, conflict-free scheduling on a given tree is an NP-hard problem. Therefore, instead of searching for an optimal schedule, the approximate solution to the subproblem of finding a minimum length schedule for a given aggregation tree is constructed using the NDR heuristic.

\subsection{Selection}
Within the selection procedure, a set of prospective parents of the next offspring is filled with solutions from the current population in the following way. Sequentially, two trees are taken from the current population in proportion to their fitness probability: the first tree of each pair is chosen randomly from the entire population, and the second tree is chosen from the remaining part of the population. Each pair should contain different trees, but the same tree may be included in many pairs. In such manner, $OffspSize$ pairs are selected.

\subsection{Crossover}
At the crossover procedure each pair of previously selected parents reproduces a child. Namely, a pair of parents $T_p^1 = (V, A_p^1)$ and $T_p^2 = (V, A_p^2)$ generates a child tree $T_c$ in the following way. Let us consider a vertex $v \in V \setminus \{s\}$ and two vertices $v_1 ,v_2 \in  V: a_1 = (v, v_1) \in A_p^1$, $a_2 = (v, v_2) \in A_p^2$. Choose an arc from $\{a_1, a_2\}$ and add it to $T_c$. If $v_1 = v_2$ then the arc $a_1$ is chosen. If adding one arc from $\{a_1, a_2\}$ to $T_c$ leads to the appearance of cycles, then another arc is chosen. In the remaining case let us introduce the weight $w_i = 1 / \delta(v_i) + 1 / |l(v) - l(v_i) - 2|$, where $\delta(v_i)$ is a degree of the vertex $v_i$ in the tree $T_p^i, i \in \{1, 2\}$. Then the arc is chosen randomly from $\{a_1, a_2\}$ with probability $P(a_i) = w_i / (w_1 + w_2), i \in \{1, 2\}$.

\subsection{Mutation}
Mutation is a randomized procedure which is applied to the tree in the current offspring. The mutation procedure is applied with probability $PM$ (a parameter of GLS) to each offspring. The mutation procedure takes as an argument (an integer parameter) $K$ --- the maximum difference (number of different arcs in the initial tree and in the modified one). This parameter is taken randomly from the interval $[1, ..., K_{max}]$, where $K_{max}$ is another parameter, inverse to its value probability (i.e., smaller modifications are more possible). The pseudocode of the mutation procedure is given in Algorithm \ref{fig:gls_m}.

\begin{algorithm}[!hbtp]
\begin{algorithmic}[1]
\STATE \emph{Input}: $G_{dir} = (V, A)$ --- a communication graph, $T = (V, A(T))$ --- a spanning tree on $G$ rooted in $s$, $K$ --- an integer parameter;
\STATE \emph{Output}: $T$ --- spanning tree in $G$ rooted in $s$;
\FORALL{$k\in\{1,...,K\}$}
\STATE Set $(v,u) \leftarrow $random arc from $A \setminus A(T)$;
\IF {$v$ is not descendant of $u$}
\STATE Remove the arc $(v, Parent(v))$ from $T$ and add the arc $(v, u)$ to $T$;
\ENDIF
\ENDFOR
\end{algorithmic}
\caption{Mutation} \label{fig:gls_m}
\end{algorithm}

\subsection{Local search}
As well as mutation, the local search procedure is applied to a subset of offspring defined by the probability $PLS$ --- another algorithm parameter. We suggest two different local search procedures. The first one, $BranchReattaching$ algorithm, is already proposed in \cite{ploErZal17}. The pseudocode of this local search procedure is presented in Algorithm \ref{fig:gls_ls_br}. At each iteration, the procedure performs a search of such arc $a = (v_1, v_2) \in A \setminus A(T)$ whose addition to $T$ (together with detaching of $v_1$ from its parent in $T$) leads to the maximum decrease of the objective function. The method $ReattachingEffect(T, v, u)$ (see Algorithm \ref{fig:ls1Effect}) calculates the change of the schedule length after detaching of $v$ from its parent in $T$ and adding an arc $(v, u)$. The whole procedure continues while the solution is improved.

\begin{algorithm}[!hbtp]
\begin{algorithmic}[1]
\STATE \emph{Input}: $G_{dir} = (V, A)$ --- a communication graph, $T = (V, A(T))$ --- a spanning tree on $G$ rooted in $s$;
\STATE \emph{Output}: $T$ --- spanning tree in $G$ rooted in $s$;
\STATE Calculate a schedule on $T$ with only the primary conflicts;
\STATE Set $improved \leftarrow true$;

\WHILE {$improved$}
\STATE Set $improved \leftarrow false$;
\STATE Set $u^* \leftarrow \emptyset$, $v^* \leftarrow \emptyset$, $bestImpr \leftarrow 0$;

\FORALL {arcs $(u, v) \in A \ A(T)$ where $u$ is not a descendant of $v$}

\STATE  Set $effect \leftarrow ReattachingEffect(T, v, u)$;

\IF {$effect < bestEffect$}
\STATE Set $u^* \leftarrow u$, $v^* \leftarrow v$, $bestEffect \leftarrow effect$, $improved \leftarrow true$;
\ENDIF

\ENDFOR

\IF {$improved$}
\STATE Remove the arc $(v^*, Parent(v^*))$ from $T$ and add the arc $(v^*, u^*)$ to $T$;
\STATE Calculate a schedule on $T$ with only the primary conflicts;
\STATE \textbf{break};
\ENDIF

\ENDWHILE
\end{algorithmic}
\caption{$BranchReattaching$ local search} \label{fig:gls_ls_br}
\end{algorithm}

\begin{algorithm}[!hbtp]
\begin{algorithmic}[1]
\STATE \emph{Input}: an initial tree $T = (A, V)$ rooted in $v_0$ with time slot assigned to each vertex; a starting vertex $v$ of an arc $(v, p)$ that is considered for inversion;
a vertex $u$ that is considered to be a new parent of $v$ in $T$.
\STATE \emph{Output}: $L(T) - L(T_1)$ --- the difference of schedule length after removing  the arc $(v, p)$ and adding the arc $(v, u)$.
\STATE Find the first common predecessor $s$ between $v$ and $u$ and two paths: $path_1$ that starts at $v$ and ends at $q$ and $path_2$ that starts at $u$ and ends at $r$ ($q$ and $r$ are both children of $s$);
\STATE Find the new minimum sending time of $q$ after removing the arc $(v, p)$ from $T$.
\IF {sending time of $q$ was not decreased} \STATE return 0. \ENDIF
\STATE Find the new minimum sending time of $u$ after adding of $v$ to its list of children.
\STATE Given new minimum sending time of $u$ find the new minimum sending time of $r$.
\STATE Given new minimum sending times of $q$ and $r$ find the new minimum sending time of $s$.
\STATE Given new minimum sending time of $s$ find find and \textbf{return} the difference of a schedule length traversing the vertices from $s$ to $v_0$.
\end{algorithmic}
\caption{$ReattachingEffect$} \label{fig:ls1Effect}
\end{algorithm}

Let us describe another local search procedure $ArcInversion$ that we also apply within GLS. As an elementary movement the sequence of the following operations can be executed within $ArcInversion$ for any vertex $v \in V$ of a current tree $T$ except its root's children: at first, the arc from $v$ to its parent $p$ is inverted; then the arc from $p$ to its parent is removed from $T$, and after that the most efficient arc that starts from $v$ and joins two obtained connected components is added to $T$. In the first steps of the algorithm and after each change of a tree, a schedule (i.e., time slots assignment) is recalculated taking into account only the primary conflicts. The pseudo code of this method is shown in Algorithm \ref{fig:ls2}.

\begin{algorithm}[!hbtp]
\begin{algorithmic}[1]
\STATE \emph{Input}: an initial tree $T = (A, V)$ rooted in $v_0$;
\STATE Calculate a schedule on $T$ with only the primary conflicts;
\STATE $improved \leftarrow true$;

\WHILE {$improved$}
\STATE $improved \leftarrow false$;

\FORALL {$v \in V \setminus \{\{v_0\} \cup \{$children of $v_0\}\}$}
\STATE Set $p^* \leftarrow Parent(v)$, $bestEffect \leftarrow 0$;

\FORALL {$u \in N(v)$ where $u$ is not a descendant of $v$ in $T$}

\STATE Set $effect \leftarrow CalcArcInversionEffect(T, v, u)$;

\IF {$effect < bestEffect$}
\STATE Set $p^* \leftarrow u$, $bestEffect \leftarrow effect$;
\ENDIF

\ENDFOR

\IF {$bestEffect < 0$}
\STATE Inverse the arc $(v, p)$, remove the arc $(p, Parent(p))$ and add the arc $(v, p^*)$ to the tree $T$;
\STATE Calculate a schedule on $T$ with only the primary conflicts;
\STATE Set $improved \leftarrow true$;
\ENDIF

\ENDFOR

\ENDWHILE
\end{algorithmic}
\caption{$ArcInversion$ local search} \label{fig:ls2}
\end{algorithm}

Similar to how it is done in $BranchReattaching$ local search procedure, in order to speed up the $ArcInversion$ local search procedure, instead of performing the modification of a tree and recalculating the schedule at each step, we calculate only the value of the schedule length change by the method $ArcInversionEffect$. This method is presented in Algorithm \ref{fig:ls2Effect}. It uses the following idea. If all minimum possible time slot values for a list of children of some vertex $v \in V$ are known, then one can easily calculate the minimum possible time slot of $v$, such that its children do not conflict with each other. If this value differs from the initial time slot of $v$, then the minimum time slot of a parent of $v$ can be updated in the same manner, and so on to the root.

Both methods $ArcInversionEffect$ and $ReattachingEffect$ in the worst case have linear complexity $O(|V|)$, but they are often performed in constant time, because they consider only 2 vertices, their neighborhood and paths in the root direction. Note, that if detachment of an arc does not decrease the length of a schedule, then the schedule length cannot be decreased by the entire local tree modification, and since only those cases when the tree is improved are taken into account inside the local search procedure, then $ReattachingEffect$ stops in line 6 (and, for the same reason, the $ArcInversionEffect$ stops in line 6).

\begin{algorithm}[!hbtp]
\begin{algorithmic}[1]
\STATE \emph{Input}: an initial tree $T = (A, V)$ rooted in $v_0$ with time slot assigned to each vertex; a starting vertex $v$ of an arc $(v, p)$ that is considered for inversion;
a vertex $u$ that is considered to be a new parent of $v$ in $T$.
\STATE \emph{Output}: $L(T) - L(T_1)$ --- the difference of schedule length after arc inversion and adding the arc $(v, u)$.
\STATE Find the first common predecessor $s$ between $v$ and $u$ and two paths: $path_1$ that starts at $v$ and ends at $q$ and $path_2$ that starts at $u$ and ends at $r$ ($q$ and $r$ are both children of $s$);
\STATE Find the new minimum sending time of $q$ after removing the arc $(v, p)$ from $T$.
\IF {sending time of $q$ was not decreased} \STATE return 0. \ENDIF
\STATE Find the new minimum sending time of $p$ after exclusion of $v$ from its list of children.
\STATE Find the new minimum sending time of $v$ after adding of $p$ to its list of children.
\STATE Find the new minimum sending time of $u$ after adding of $v$ to its list of children.
\STATE Given new minimum sending time of $u$ find the new minimum sending time of $r$.
\STATE Given new minimum sending times of $q$ and $r$ find the new minimum sending time of $s$.
\STATE Given new minimum sending time of $s$ find find and \textbf{return} the difference of a schedule length traversing the vertices from $s$ to $v_0$.
\end{algorithmic}
\caption{$ArcInversionEffect$} \label{fig:ls2Effect}
\end{algorithm}

\subsection{Join}
At the join procedure $PopSize$ solutions from the current population and the current offspring, which have the largest fitness values, are chosen to fill the population of the next generation.

\section{VNS based heuristic}\label{sec:heuristics}
In this section, we suggest a new approach where different aggregation trees are examined within an algorithm based on the variable neighborhood search (VNS) metaheuristic. As well as it was done within the GLS algorithm, here we apply different local search procedures for the relaxed problem, where only the primary conflicts are taken into account. Each time, after applying local search procedure, some heuristic algorithm is executed to calculate a schedule (and, maybe, slightly modify a tree) taking all conflicts into account. Also, an algorithm contains a shaking operator which constructs a new tree at random in some predefined neighborhood of the current tree, and the maximum remoteness (i.e., the number of arcs that belong to one tree and don't belong to another) of the obtained tree from the initial tree is bounded by a given integer parameter.

The pseudo code of the proposed VNS based algorithm is presented in Algorithm \ref{fig:vns}. The names $Shaking$, $LS_1, ..., LS_{l_{\max}}$ and $Schedule$, correspondingly, denote the shaking operator, the local search methods taking into account only the primary conflicts, and the approximate scheduling algorithm for the entire problem. Note that some of the known scheduling methods (e.g., NDR) may modify the aggregation tree, the final tree is returned by the scheduling operator together with a schedule length.

\begin{algorithm}[!hbtp]
\begin{algorithmic}[1]
\STATE \emph{Input}: an initial tree $T$; a procedure $Shaking(T, K)$; $K_{\max} > 0$; a set of local search algorithms $LS_1(T), ..., LS_{l_{\max}}(T)$ for the problem with only the primary conflicts; a scheduling operator $Schedule(T)$.
\STATE Apply scheduling operator: $(T, L) \leftarrow Schedule(T)$;
\WHILE {the stopping criteria is not met}
\STATE $K \leftarrow 0$;
\WHILE {$K \leq K_{\max}$}
\STATE Perform shaking operator: $T^{\prime} \leftarrow Shaking(T, K)$;
\STATE Apply scheduling operator: $(T^{\prime}, L^{\prime}) \leftarrow Schedule(T^{\prime})$;
\WHILE {$l \leq l_{\max}$}
\STATE Perform local search: $T^{\prime\prime} \leftarrow LS_l(T^{\prime})$;
\STATE $(T^{\prime\prime}, L^{\prime\prime}) \leftarrow Schedule(T^{\prime\prime})$;
\IF {$L^{\prime\prime} <  L^{\prime}$}
\STATE $T^{\prime} \leftarrow T^{\prime\prime}$; $L^{\prime} \leftarrow L^{\prime\prime}$; $l \leftarrow 1$;
\ELSE
\STATE $l \leftarrow l+1$;
\ENDIF
\ENDWHILE
\IF {$L^{\prime} <  L$}
\STATE $T \leftarrow T^{\prime}$; $L \leftarrow L^{\prime}$; $K \leftarrow 1$;
\ELSE
\STATE $k \leftarrow k+1$;
\ENDIF
\ENDWHILE
\ENDWHILE
\end{algorithmic}
\caption{VNS-based algorithm} \label{fig:vns}
\end{algorithm}

For the local search, we use two heuristics, $BranchReattaching$ and $ArcInversing$, that have already been described in the previous section. In the shaking phase, we apply the $Mutation$ algorithm from the previous section, that has the maximum remoteness of the shaken solution from the initial one as an input parameter $K$. As an approximate algorithm of calculation of a conflict-free schedule on a given aggregation tree we used the NDR heuristic \cite{Pan16}.

\section{Simulation}
In order to verify the efficiency of the proposed algorithms we have implemented them in C++ programming language. We have also implemented the best state-of-the-art methods MLST \cite{Pan16} and RH \cite{Beier00} to generate an aggregation tree, and NDR \cite{Pan16} to construct a schedule on the aggregation tree. The ILP formulation from \cite{Tian11} was used to obtain optimal solutions for small-sized test instances using the CPLEX package.

For generating the test instances we used the examples of points allocation for the Euclidian Steiner Tree Problem from the Beasley's OR-Library (http://people.brunel.ac.uk/~mastjjb/jeb/orlib). In each of these instances, a set of points are spread inside a planar square with a side of unit length. To simulate the communication network, we generated a unit disk graph (UDG) based topology on the given set of nodes. We use the following definition of UDG taken from \cite{Clark90}. Given $n$ points in the plane and some specified bound $d$, the \emph{unit disk graph} is undirected graph with $n$ vertices corresponding to the given $n$ points, where an edge connects two vertices if and only if the Euclidean distance between the two corresponding points does not exceed the \emph{critical distance} $d$. This is one of the most commonly used models for the communication networks, where each element has the same transmission range that is equal to the critical distance of the corresponding UDG. As a sink node we always chose the vertex that lies closer to the square center.

We have tested the different pairs of ($n$, $d$), where $n = 10, 20, 100, 250, 500, 1000$, and $d \leq 0.5$. Obviously, in cases of small density (when $d$ is too small for a given $n$) the UDG may be disconnected. Therefore, we left only those cases where the connectivity condition holds.

During the experiment, we tested the following heuristics for generating an aggregation tree: Dijkstra algorithm (Dij), since the shortest-path trees appear to be rather efficient in low-dense cases, Round Heuristic (RH) \cite{Beier00}, that appears to be very efficient for the simplified problem with only the primary conflicts, and MLST \cite{Pan16}, that enables a small schedule length lower bound and, according to the conclusions in \cite{Pan16}, outperforms many other known heuristics of the aggregation tree construction. As it was described above, all of these trees are included into the initial population of genetic algorithm. As a start solution of VNS we took the best of these trees (i.e., we applied NDR to construct a schedule on each tree and compares the schedule lengths). The following heuristic algorithms were compared:

\begin{itemize}
  \item Generation of aggregation tree and scheduling:
  \begin{itemize}
    \item H1 (MLST and NDR);
    \item H2 (RH and NDR);
    \item H3 (Dij and NDR);
  \end{itemize}
  \item Metaheuristic algorithms:
  \begin{itemize}
    \item GLS1 (GLS with $ArcInversing$ local search);
    \item GLS2 (GLS with $BranchReattaching$ local search);
    \item VNS.
  \end{itemize}
\end{itemize}

For the GA-based algorithms the following parameters empirically were chosen to get the best results: $PopSize$ = 50, $OffspSize$ = 20, $FPItCount$ = 150, $SPProportion$ = 0.6, $PM$ = 0.5, $PLS$ = 0.5, $K_{max} = \lfloor n/3 \rfloor$. The the VNS based algorithms we chose $K_{max} = 30$ because in most cases this value allowed obtaining the best solutions in a shorter time period. As a stopping condition of both metaheuristics the next rule was taken: an algorithm proceeds until its incumbent remains the same 3 times in a row.

The results of the experiment are presented in Tables \ref{tab:sdim} and \ref{tab:ldim}. In both tables the first 3 columns contain the test instance properties: the number of vertices, $n$; the distance upper bound of a unit disk graph, $d$; the instance case number in the OR Library, nr. On each test instance we launched GLS1, GLS2, and VNS 20 times and calculated the average values of the obtained schedule length (SL.av). SL stands for the schedule lengths obtained by other algorithms, that were launched once on each tested instance. The best values among all algorithms are marked bold in both tables.

In Table \ref{tab:sdim} the results for small size samples are presented. In these cases, an optimal solution was obtained by CPLEX. For each metaheuristic based algorithm and for each test instance the percentage of runs when the obtained solution is optimal is given in the column Opt. The last row, shows the overall percentages of the cases when the obtained solution is optimal. It can be noticed that GLS and VNS based heuristics found the optimal solution in more than 90\% of all cases, while the performance of the other heuristics is significantly worse. Besides, although it is not shown in the table, each time the GLS and VNS based heuristics constructed a solution, whose schedule length exceeded the optimal schedule length by not more than 1.

\begin{table}[!hbtp]
\resizebox{0.85\textwidth}{!}
 {
 \begin{minipage}{\textwidth}
\begin{tabular}{|c|c|c|c|c|c|c|c|c|c|c|c|c|}
\hline
\multirow{2}{*}{$n$} &\multirow{2}{*}{$d$} &\multirow{2}{*}{nr} &CPLEX &H1 &H2 &H3 &\multicolumn{2}{c|}{GLS1} &\multicolumn{2}{c|}{GLS2} &\multicolumn{2}{c|}{VNS} \\
\cline{4-13}
%& & &\rotatebox{90}{RH\_NDR} &\rotatebox{90}{GA\_AI} &\rotatebox{90}{GA\_BR} &\rotatebox{90}{VNS} &\rotatebox{90}{RH\_NDR} &\rotatebox{90}{GA\_AI} &\rotatebox{90}{GA\_BR} &\rotatebox{90}{VNS} \\
& & &SL &SL &SL &SL &SL.av &Opt (\%) &SL.av &Opt (\%) &SL.av &Opt (\%)\\
\hline

10	&0.5	&1	&5	&7	&7	&6	&\textbf{5}	&100	&5.05	&95	&5.05	&95\\
10	&0.5	&2	&5	&6	&\textbf{5}	&7	&\textbf{5}	&100	&\textbf{5}	&100	&\textbf{5}	&100\\
10	&0.5	&3	&5	&7	&6	&7	&\textbf{5.1}	&90	&5.15	&85	&\textbf{5.1}	&90\\
10	&0.5	&4	&5	&6	&\textbf{5}	&\textbf{5}	&\textbf{5}	&100	&\textbf{5}	&100	&\textbf{5}	&100\\
10	&0.5	&5	&5	&7	&6	&7	&\textbf{5}	&100	&\textbf{5}	&100	&\textbf{5}	&100\\
10	&0.5	&6	&5	&7	&\textbf{5}	&6	&\textbf{5}	&100	&\textbf{5}	&100	&\textbf{5}	&100\\
10	&0.5	&8	&5	&\textbf{5}	&\textbf{5}	&\textbf{5}	&\textbf{5}	&100	&\textbf{5}	&100	&\textbf{5}	&100\\
10	&0.5	&9	&5	&6	&7	&9	&\textbf{5.6}	&40	&5.65	&35	&6	&0\\
10	&0.5	&10	&5	&6	&6	&6	&5.45	&55	&\textbf{5.4}	&60	&5.45	&55\\
10	&0.5	&11	&5	&6	&6	&6	&\textbf{5}	&100	&\textbf{5}	&100	&\textbf{5}	&100\\
10	&0.5	&12	&6	&\textbf{6}	&7	&8	&\textbf{6}	&100	&\textbf{6}	&100	&\textbf{6}	&100\\
10	&0.5	&13	&6	&7	&8	&9	&\textbf{6}	&100	&\textbf{6}	&100	&\textbf{6}	&100\\
10	&0.5	&14	&6	&\textbf{6}	&\textbf{6}	&\textbf{6}	&\textbf{6}	&100	&\textbf{6}	&100	&\textbf{6}	&100\\
10	&0.5	&15	&6	&7	&\textbf{6}	&7	&\textbf{6}	&100	&\textbf{6}	&100	&\textbf{6}	&100\\
20	&0.33	&6	&7	&\textbf{7}	&\textbf{7}	&\textbf{7}	&\textbf{7}	&100	&\textbf{7}	&100	&\textbf{7}	&100\\
20	&0.33	&8	&6	&8	&7	&7	&\textbf{6}	&100	&6.3	&70	&\textbf{6}	&100\\
20	&0.33	&13	&7	&9	&\textbf{7}	&8	&\textbf{7}	&100	&\textbf{7}	&100	&\textbf{7}	&100\\
\hline
\multicolumn{4}{|c|} {Opt (\%)} &20 	&40  &20 &\multicolumn{2}{c|}{93.24} &\multicolumn{2}{c|}{90.88}   &	\multicolumn{2}{c|}{90.5}\\
\hline

\end{tabular}
\medskip
\end{minipage}}
\caption{Schedule lengths obtained by different algorithms in small size cases.}\label{tab:sdim}
\end{table}

\begin{table}[!hbtp]
\resizebox{0.7\textwidth}{!}
 {
 \begin{minipage}{\textwidth}
\begin{tabular}{|c|c|c|ccc|ccc|ccc|ccc|ccc|}
\hline

\multirow{2}{*}[-1em]{$n$} &\multirow{2}{*}[-1em]{$d$} &\multirow{2}{*}[-1em]{nr} &\multicolumn{3}{c|}{SL} &\multicolumn{3}{c|}{SL.best} &\multicolumn{3}{c|}{SL.av} &\multicolumn{3}{c|}{SL.sd} &\multicolumn{3}{c|}{Time (in sec.)}\\

\cline{4-18}

& & &\multirow{1}{*}[0.75em]{H1} &\multirow{1}{*}[0.75em]{H2}  &\multirow{1}{*}[0.75em]{H3}
&\rotatebox{90}{GLS1} &\rotatebox{90}{GLS2} &\rotatebox{90}{VNS}
&\rotatebox{90}{GLS1} &\rotatebox{90}{GLS2} &\rotatebox{90}{VNS}
&\rotatebox{90}{GLS1} &\rotatebox{90}{GLS2} &\rotatebox{90}{VNS}
&\rotatebox{90}{GLS1} &\rotatebox{90}{GLS2} &\rotatebox{90}{VNS}\\

\hline

\multirow{13}{*}{100}	&\multirow{5}{*}{0.3}
    &1  &22	&19	&26	&19	&19	&\textbf{18}	&19	    &19 	&\textbf{18.8}	&0	    &0	    &0.41	&2.04	&\textbf{2.27}	&2.57\\
&	&2	&24	&23	&35	&19	&19	&\textbf{18}	&20.15	&20.4	&\textbf{19.3}	&0.49	&0.68	&0.57	&4.53	&\textbf{4.36}	&4.54\\
&	&3	&23	&21	&25	&19	&\textbf{18}	&\textbf{18}	&19.45	&19.25	&\textbf{19.15}	&0.6	&0.55	&0.49	&\textbf{3.25}	&3.74	&3.95\\
&	&4	&24	&22	&32	&19	&\textbf{18}	&\textbf{18}	&19.05	&18.95	&\textbf{18.8}	&0.22	&0.22	&0.41	&3.27	&3.84	&\textbf{2.97}\\
&	&5	&22	&19	&26	&\textbf{18}	&\textbf{18}	&\textbf{18}	&18.3	&18.4	&\textbf{18}	    &0.47	&0.5	&0	    &2.57	&2.76	&\textbf{2.24}\\
\cline{2-18}
&\multirow{5}{*}{0.4}
    &1	&28	&26	&49	&\textbf{24}	&\textbf{24}	&\textbf{24}	&25.6	&25.45	&\textbf{24.9}5	&0.6	&0.69	&0.22	&5.73	&7.4	&\textbf{4.07}\\
&	&2	&28	&28	&56	&\textbf{25}	&\textbf{25}	&\textbf{25}	&26.45	&\textbf{25.95}	&26.65	&0.76	&0.94	&0.67	&\textbf{7.97}	&11.48	&10.07\\
&	&3	&26	&26	&42	&\textbf{23}	&\textbf{23}	&\textbf{23}	&24.5	&\textbf{24.25}	&24.7	&1.1	&0.97	&0.8	&\textbf{6.19}	&7.85	&7.79\\
&	&4	&30	&28	&52	&25	&25	&\textbf{24}	&25.55	&25.85	&\textbf{25.15}	&0.69	&0.81	&0.59	&\textbf{7.97}	&9.87	&8.36\\
&	&5	&29	&27	&48	&24	&24	&\textbf{23}	&25.5	&25.2	&\textbf{24.55}	&0.69	&0.7	&0.6	&\textbf{7.14}	&9.02	&7.65\\
\cline{2-18}
&\multirow{3}{*}{0.5}
    &1	&38	&35	&76	&\textbf{32}	&34	&\textbf{32}	&34.35	&34.8	&\textbf{33.25}	&0.81	&0.41	&0.64	&\textbf{11.27}	&17.51	&11.47\\
&	&2	&39	&36	&82	&33	&\textbf{31}	&32	&35    	&33.85	&\textbf{32.75}	&1.17	&1.35	&0.44	&13.79	&20.79	&\textbf{10.3}\\
&	&3	&34	&34	&74	&30	&\textbf{29}	&31	&31.45	&\textbf{31.35}	&32.3	&0.76	&1.09	&0.8	&\textbf{12.91}	&18.85	&13.49\\

\hline

\multirow{13}{*}{250}	&\multirow{5}{*}{0.2}
	&1	&31	&28	&44	&\textbf{24}	&\textbf{24}	&\textbf{24}	&25.6	&25.5	&\textbf{24.7}	&0.75	&0.61	&0.57	&20.71	&19.95	&\textbf{17.63}\\
&	&2	&33	&25	&32	&24	&\textbf{23}	&\textbf{23}	&24.3	&24.85	&\textbf{23.35}	&0.47	&0.49	&0.49	&10.49	&\textbf{8.19}	&12.62\\
&	&3	&34	&27	&39	&\textbf{24}	&\textbf{24}	&\textbf{24}	&25.45	&25.5	&\textbf{24.85}	&0.83	&1.05	&0.37	&17.04	&16.61	&\textbf{13.73}\\
&	&4	&30	&25	&32	&\textbf{23}	&24	&\textbf{23}	&24.3	&24.65	&\textbf{23.05}	&0.73	&0.49	&0.22	&10.57	&\textbf{8.7}	&11.27\\
&	&5	&32	&24	&36	&24	&24	&\textbf{23}	&24    	&24    	&\textbf{23.6}	&0  	&0     	&0.5	&\textbf{9.73}	&10.07	&13.88\\
\cline{2-18}
&\multirow{5}{*}{0.3}
	&1	&49	&42	&79	&37	&37	&\textbf{36}	&39.05	&39.6	&\textbf{37.05}	&1.32	&1.9	&0.6	&59.73	&69.13	&\textbf{40.41}\\
&	&2	&48	&40	&52	&39	&39	&\textbf{37}	&39.9	&39.75	&\textbf{38.7}	&0.31	&0.44	&0.73	&39.55	&44.03	&\textbf{36.88}\\
&	&3	&53	&43	&74	&\textbf{36}	&\textbf{36}	&\textbf{36}	&38.15	&38.05	&\textbf{36.85}	&0.99	&1.0	&0.59	&59.49	&69.71	&\textbf{38.65}\\
&	&4	&47	&41	&61	&\textbf{36}	&\textbf{36}	&37	&37.7	&\textbf{37.6}	&38.1	&1.03	&0.75	&0.64	&64.96	&62.06	&\textbf{32.78}\\
&	&5	&49	&42	&76	&\textbf{37}	&\textbf{37}	&\textbf{37}	&38.05	&38.6	&\textbf{37.6}	&0.89	&0.88	&0.5	&66.26	&68.41	&\textbf{32.85}\\
\cline{2-18}
&\multirow{3}{*}{0.4}
	&1	&70	&59	&121	&\textbf{52}	&\textbf{52}	&53	&54.45	&54.05	&\textbf{53.9}	&1.32	&1.23	&0.91	&151.62	&190.65	&\textbf{83.21}\\
&	&2	&75	&64	&118	&\textbf{55}	&\textbf{55}	&\textbf{55}	&57.45	&57.75	&\textbf{56.9}	&1.23	&1.41	&0.91	&163.06	&187.75	&\textbf{66.66}\\
&	&3	&64	&62	&129	&\textbf{53}	&\textbf{53}	&54	&55.15	&54.85	&\textbf{55.1}	&1.09	&0.99	&0.72	&180.53	&200.21	&\textbf{84.76}\\

\hline

\multirow{9}{*}{500}	&\multirow{5}{*}{0.1}
	&1	&29	&24	&25	&\textbf{21} &\textbf{21} &\textbf{21}	&22.65	&22.85	&\textbf{21.95}	&0.67	&0.99	&0.39	&25.92	&24.77	&\textbf{19.75}\\
&	&2	&28	&25	&24	&22	&\textbf{21} &\textbf{21} &22.2	&22.1	&\textbf{21.95}	&0.41	&0.45	&0.22	&21.3	&17.81	&\textbf{14.43}\\
&	&3	&25	&23	&24	&\textbf{21} &\textbf{21} &\textbf{21} &21.75	&21.8	&\textbf{21.1}	&0.44	&0.41	&0.31	&16.96	&\textbf{15.4}	&16.96\\
&	&4	&28	&24	&25	&\textbf{21} &\textbf{21} &\textbf{21} &22	    &22.25	&\textbf{21.35}	&0.56	&0.55	&0.49	&21.7	&19.45	&\textbf{15.43}\\
&	&5	&25	&22	&23	&\textbf{21} &\textbf{21} &\textbf{21} &21.95	&21.9	&\textbf{21}  	&0.22	&0.31	&0	    &12.31	&\textbf{11.42}	&14.16\\
\cline{2-18}
&\multirow{3}{*}{0.2}
	&1	&51	&41	&68	&40	&40	&\textbf{38}	&41.6	&41.3	&\textbf{39.6}	&0.75	&0.92	&0.6	&99.66	&104.24	&\textbf{82.18}\\
&	&2	&49	&41	&68	&41	&41	&\textbf{40}	&41	    &41	    &\textbf{40.5}	&0	    &0	    &0.51	&88.86	&85.15	&\textbf{73.27}\\
&	&3	&49	&\textbf{39}	&66	&\textbf{39}	&\textbf{39}	&\textbf{39}	&\textbf{39}	    &\textbf{39}	    &\textbf{39}	&0	    &0	    &0	&82.79	&\textbf{80.77}	&96.92\\
\cline{2-18}
&\multirow{1}{*}{0.3}
	&1	&85	&73	&144	&65	&65	&\textbf{64}	&67.35	&66.95	&\textbf{65.9}	&1.31	&1.67	&0.79	&571.87	&667.17	&\textbf{241.35}\\

\hline

\multirow{3}{*}{1000}	&\multirow{2}{*}{0.1}
	&1	&42	&33	&38	&\textbf{30}	&\textbf{30}	&\textbf{30}	&31.45	&31.65	&\textbf{30.25}	&0.83	&0.88	&0.44	&205.29	&169.85	&\textbf{113.12}\\
&	&2	&37	&35	&39	&\textbf{30}	&31	&\textbf{30}	&31.4	&31.6	&\textbf{30}	&0.82	&0.82	&0	&163.23	&155.63	&\textbf{97.35}\\
\cline{2-18}
&\multirow{1}{*}{0.2}
	&1	&89	&70	&123	&68	&\textbf{67}	&\textbf{67}	&70.30	&69.05	&\textbf{68.25}	&1.38	&1.00	&0.64	&1196.24	&1068.21	&\textbf{462.78}\\

\hline

\end{tabular}
\medskip
\end{minipage}}
\caption{Schedule lengths obtained by different algorithms in large size cases.}\label{tab:ldim}
\end{table}

The results for larger size cases are presented in Table \ref{tab:ldim}. In addition to the average schedule lengths, this table contains standard deviations (SL.sd), schedule lengths on best solutions (SL.best), and the average calculation times for the algorithms GLS1, GLS2, and VNS. For each tested combination of $n$ and $d$, we took 5 (or, in some cases, less) first examples from the Beasley's OR-Library. As it is seen from the table, in almost all cases, the fast heuristics (H1, H2, and H3) failed to construct an optimal solution, since in each case, except the instance $n = 500, d = 0.2, $ nr $ = 3$, one of our algorithms GLS1, GLS2, or VNS found a solution with shorter schedule length. Concerning the fast heuristics, H2, which is based on the RH tree construction method, appeared to be significantly more efficient than other algorithms, and, as it was expected, the efficiency of H3, where the shortest-path tree is used, falls down with increase of the communication graph density (which depends on $d$). In most of the cases, VNS based heuristic appeared to be more efficient than any of two GLS based heuristics as in terms of average schedule length, so in terms average running time. If we compare the schedule lengths on the best solutions obtained by fast heuristics and the schedule lengths on the best solutions obtained by metaheuristic based approaches, then significant superiority of the last mentioned ones can be noticed. Thus, this difference exceeds 1 in 84.2\% of cases, it exceeds 2 in 73.7\% of cases, it exceeds 3 in 42.1\% of cases, and it equals 9 in 3 cases. It can be noticed that this gap grows when $d$ increases. This can be explained by an assumption, that the strategy of solving the problem in two separate stages (the single aggregation tree building followed by the scheduling on it) does not allow to find near-optimal solutions when the graph density is large, because in these cases there are too many conflicts that should be taken care of in a scheduling stage, and it becomes harder to guess an appropriate aggregation tree before scheduling.

\begin{figure}[!hbtp]
\centering
\subfloat[\label{fig:ex_commGraph}Communication graph. $|E| = 1200$]{\includegraphics[width=0.33\textwidth]{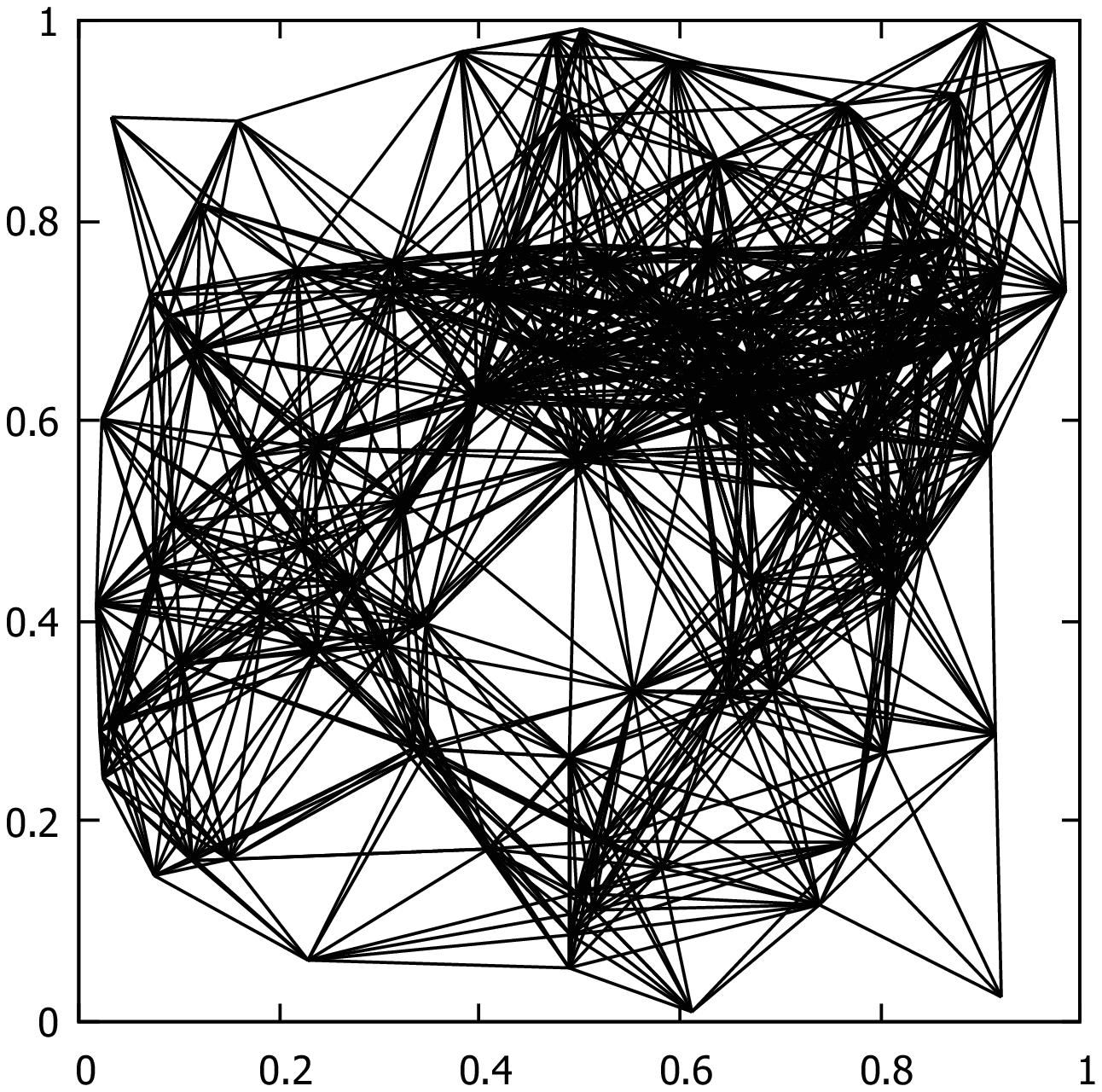}} \hfill
\subfloat[\label{fig:ex_H1}H1. SL=24]{\includegraphics[width=0.33\textwidth]{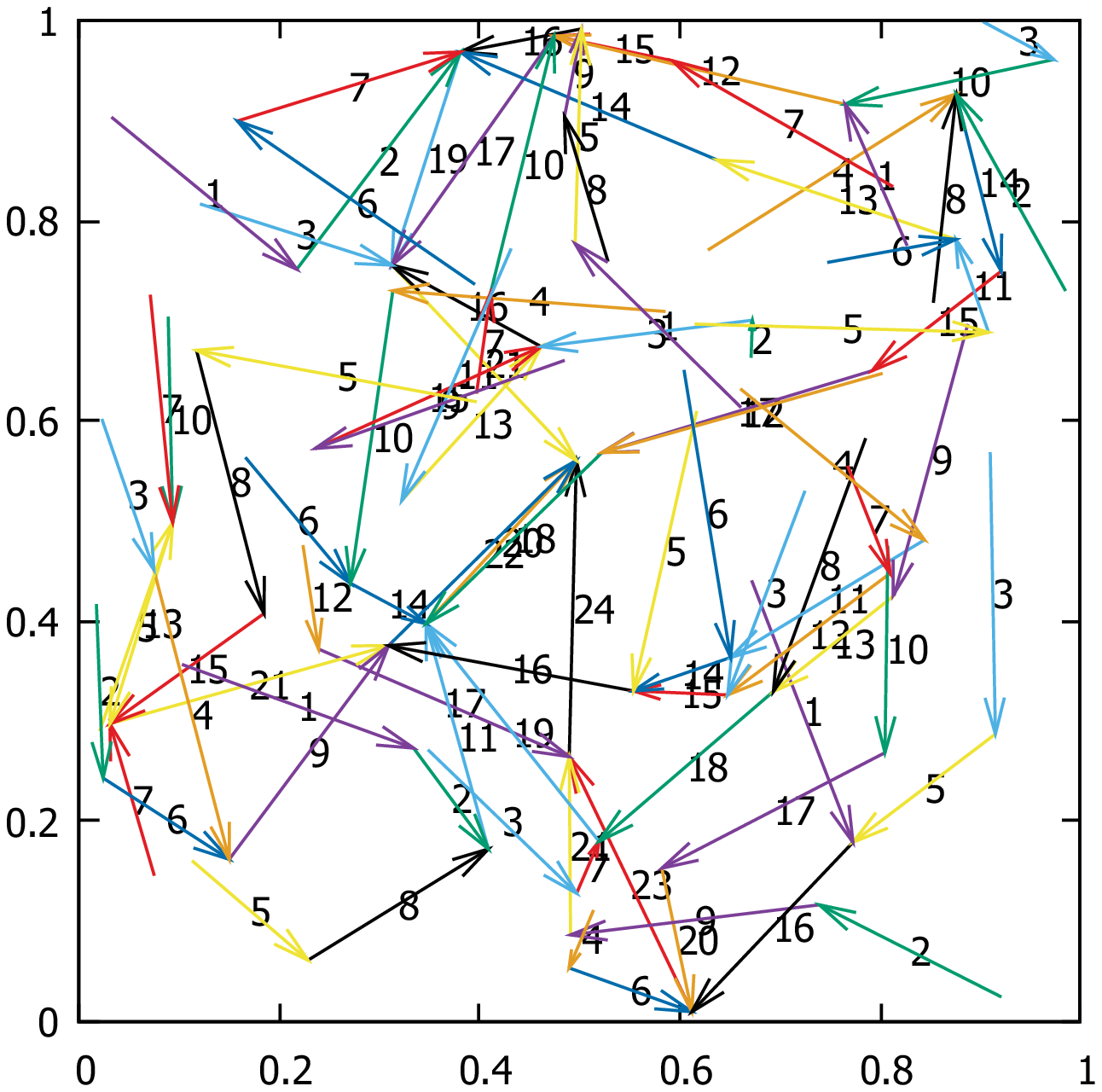}} \hfill
\subfloat[\label{fig:ex_H2}H2. SL=23]{\includegraphics[width=0.33\textwidth]{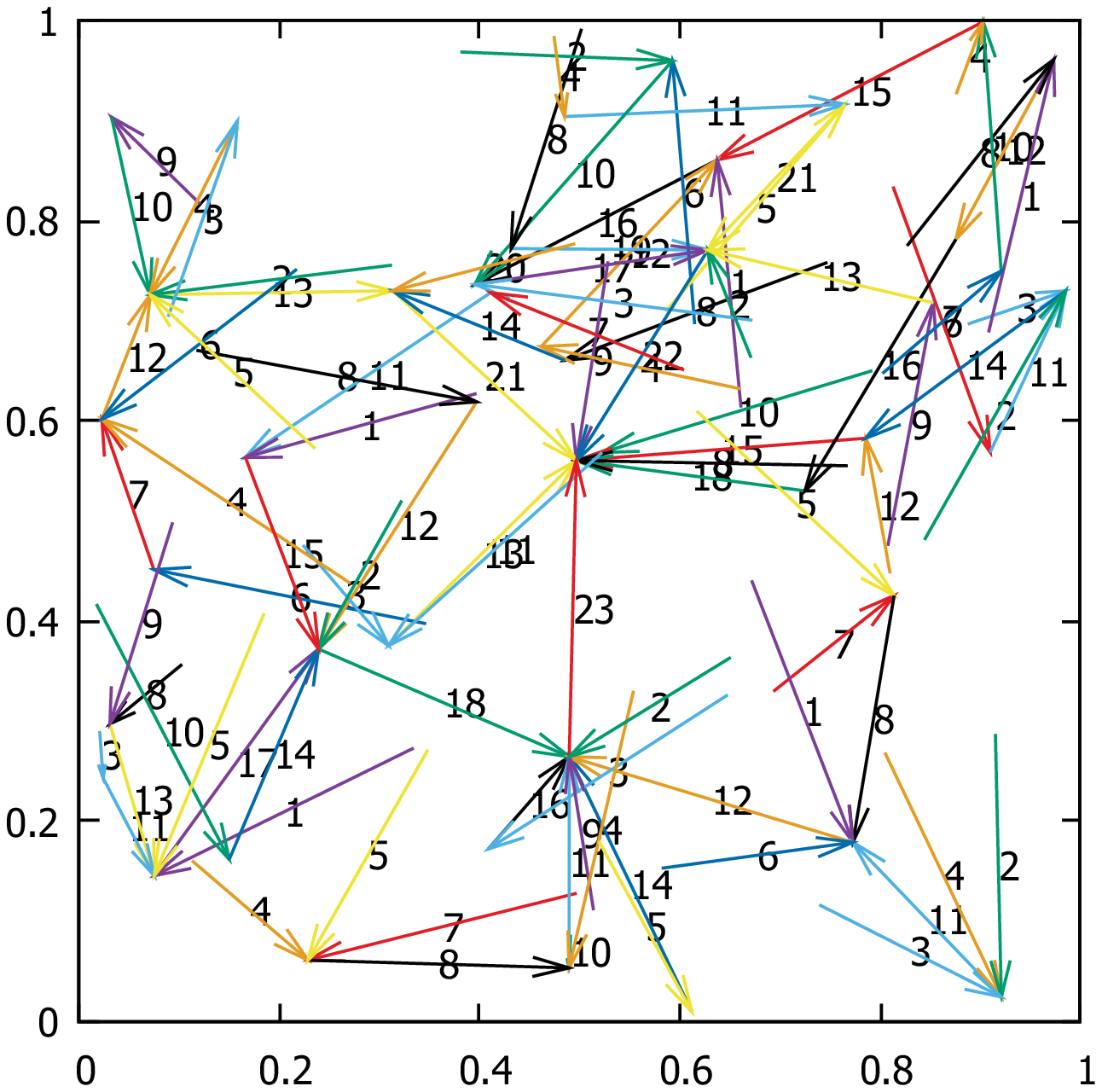}} \hfill
\subfloat[\label{fig:ex_H3}H3. SL=35]{\includegraphics[width=0.33\textwidth]{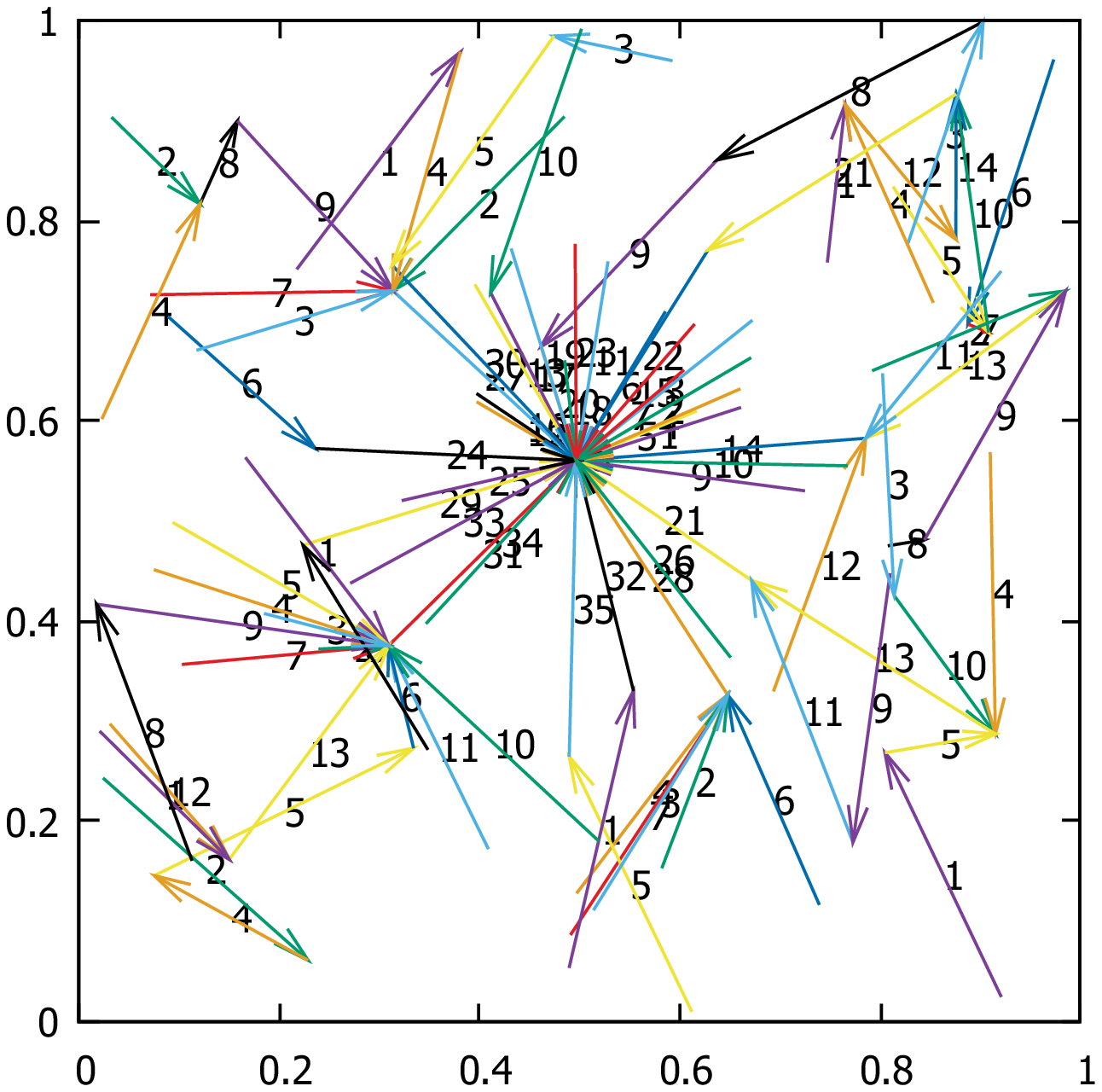}} \hfill
\subfloat[\label{fig:ex_GLS1}GLS1. SL=19]{\includegraphics[width=0.33\textwidth]{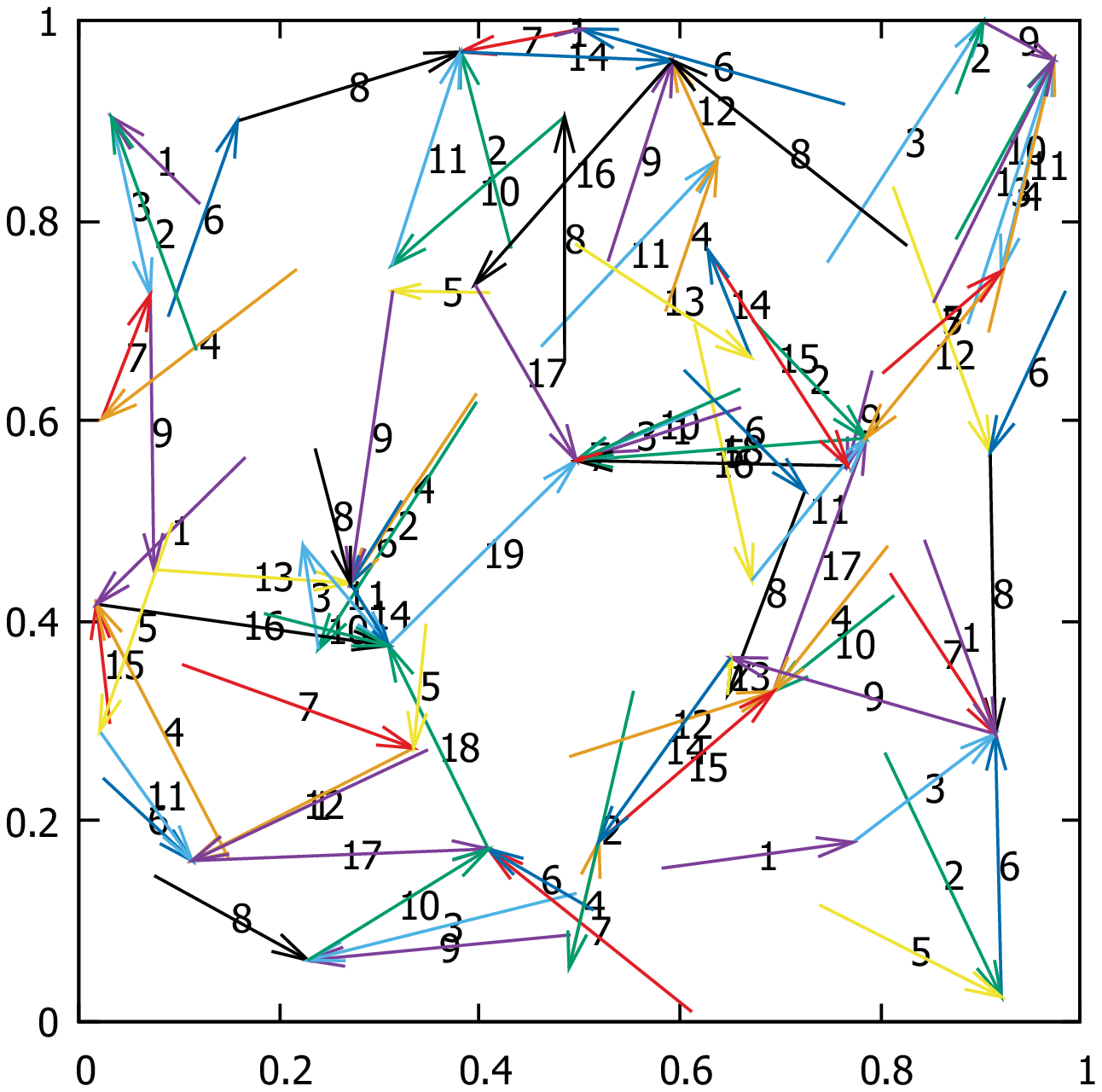}} \hfill
\subfloat[\label{fig:ex_GLS2}GLS2. SL=19]{\includegraphics[width=0.33\textwidth]{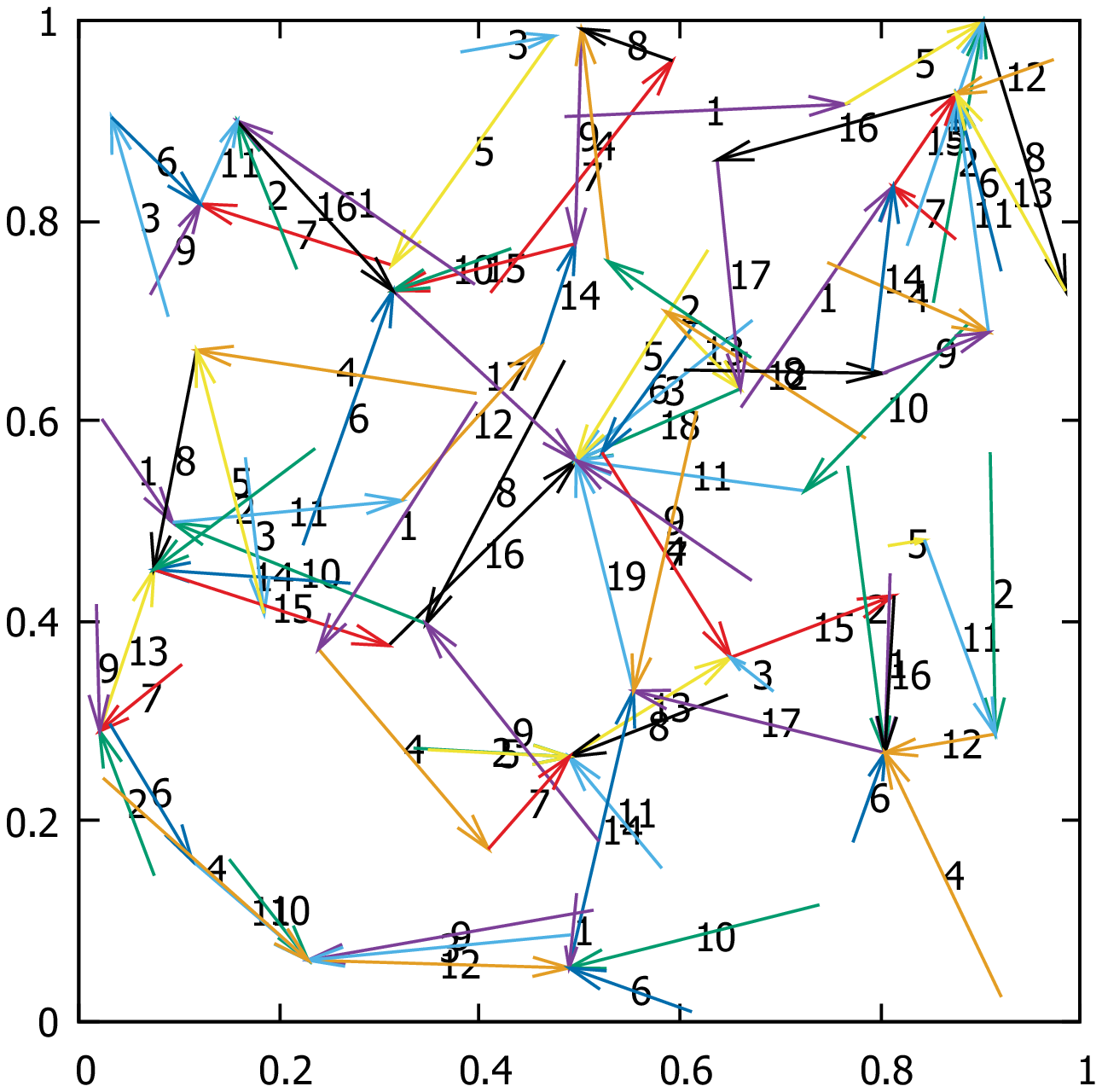}} \hfill
\subfloat[\label{fig:ex_VNS}VNS. SL=18]{\includegraphics[width=0.33\textwidth]{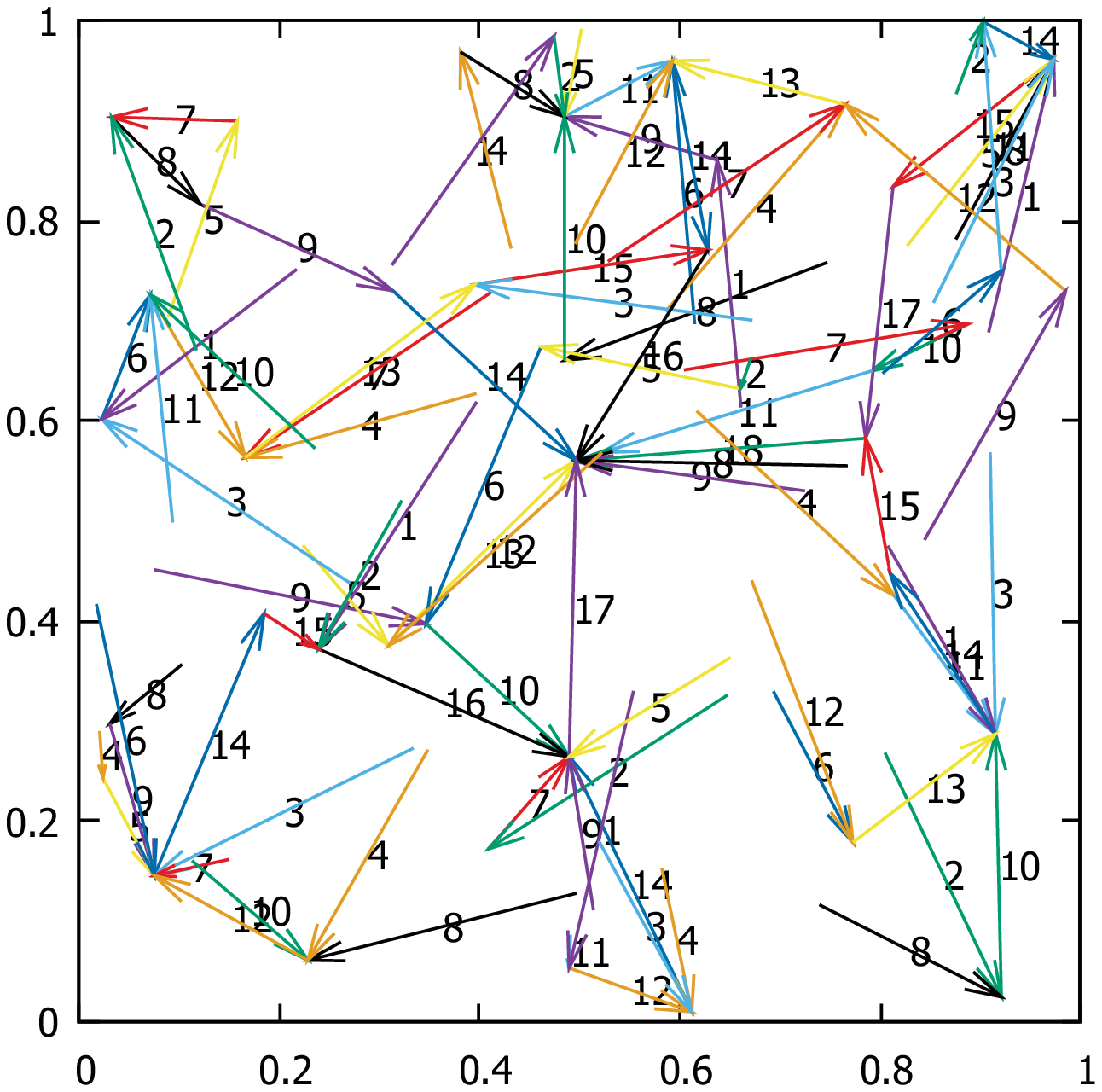}} \hfill
\caption{The unit disk graph with $n$ = 100 and $d$ = 0.2 on the point set case nr. 2 of the Beasley's OR-Library and the solutions obtained by different algorithms on this case.} \label{fig:ex}
\end{figure}

As an illustration, in Fig. \ref{fig:ex} the communication graph and the solutions obtained by different algorithms in the case $n=100, d = 0.2, $ nr $ = 2$ are presented. The numbers on the arcs denote the time slots when the messages are transmitted along the arcs, and the arcs with the same transmission time slot are equally colored. Note that the aggregation trees constructed by the metaheuristic based algorithms are very different from the trees obtained by H1, H2, and H3, that were used as their initial solutions.

\section{Conclusion}

In this paper, we considered a problem of conflict-free minimum length aggregation scheduling in a wireless sensor network. This is known NP-hard problem. Moreover, even a scheduling problem on a given aggregation tree remains NP-hard.  Therefore, the development of efficient approximate algorithms is very essential for this problem. One of the most common approaches to solve this problem approximately consists in two stages: constructing an appropriate aggregation tree at the first stage and finding a near-optimal schedule on this tree at the second stage. We noticed that such approach has the following disadvantage. Although the corresponding algorithms are often rather fast, the obtained solution could be bad, because, in such methods, the aggregation tree is chosen once before scheduling, and remains fixed (or almost fixed, as in NDR). We propose two new heuristic algorithms that have no the mentioned flaw, since they consider different aggregation trees, obtained by some known constructive heuristic or by modification of other trees. Our methods are based on two metaheuristics: genetic local search (or memetic algorithm) and variable neighborhood search. They use two variants of local search procedure. The first one, $BranchReattaching$, is already proposed in \cite{ploErZal17}, and the second, $ArcInversing$, is new. During the numerical experiment, we established the best combinations of parameters of our algorithms and then compared them with other known approaches. The extensive simulation has shown that, on average, the both of our algorithms outperform the best of the known approaches. On average, in the majority of the tested cases, VNS appeared to be more efficient than GLS both in terms of schedule length and running time, and its superiority becomes more noticeable when the number of sensors and the transmission distance increase.

\section*{Disclosure statement}

No potential conflict of interest was reported by the authors.

\section*{Funding}

The research is supported by the Russian Science Foundation (project 18-71-00084).

%
% ---- Bibliography ----
%
% BibTeX users should specify bibliography style 'splncs04'.
% References will then be sorted and formatted in the correct style.
%
% \bibliographystyle{splncs04}
% \bibliography{mybibliography}
%

\end{document}